\documentclass [12pt] {article}

\begin{document}

\begin{titlepage}
\begin{center}

{\Large \bf Megaton Modular Multi-Purpose Neutrino Detector}

\vspace{1.5in}

{\bf Alfred K. Mann }\\
{\bf Department of Physics \& Astronomy}\\
{\bf University of Pennsylvania}\\
{\bf Philadelphia, PA 19104}\\
\vspace{0.1in}
{\bf  July 24, 2001\\}
\vspace{1.5in}
\end{center}
\begin{abstract}\
This brief note outlines a detector for the Homestake Laboratory 
and the physics experiments that might be done with it. The note 
might go on record as a letter of intent to be converted to a 
formal proposal at a later time.

\end{abstract}

\end{titlepage}
\newpage

A major challenge to be faced in the construction of a multipurpose
underground laboratory lies in the design of the rooms required by the
most massive detectors, their sizes, implementation, configuration,
and depth. Smaller rooms present other, usually less serious
problems. Accordingly, it is useful to address the physics that 
demands massive detectors at an early time so that the issues
and expense involved in constructing them are appropriately delineated
in the overall description of the laboratory. That is the purpose 
of this note.

The characteristics of the Homestake Underground Laboratory, 
very deep laboratory
chambers and very strong rock, open  the possibility of a new 
generation of
extremely large detectors that can explore hitherto unexamined 
domains of the Universe
and of fundamental physics interactions. Among the recognized 
scientific goals of such a detector system are:
\begin{enumerate}
\item[] a.  Search for proton decays one or more orders of magnitude 
in lifetime
beyond the
present experimental limits
\item[] b.  Search for localized or diffuse astrophysical sources of 
ultra high
energy (UHE)
neutrinos.
\item[] c.  Extend the investigation of neutrino flavor transitions of
atmospheric
neutrinos beyond the range explored by Superkamiokande by significantly 
more statistics.
\item[] d.  Detect neutrino bursts from supernova and, especially, 
detect the neutrinos from the electron capture reaction at the 
formation of the protoneutron star.
\item[] e.  Provide a potential target for future  very intense 
terrestrial
sources of neutrinos with emphasis on a search for CP invariance 
violation in the neutrino sector.
\item[] f.  Permit additional high statistics investigations of 
solar neutrino
emission,
if that is still required.
\end{enumerate}

The large water Cerenkov detectors at Lake Erie, Sudbury and Kamioka,
(Kamiokande~II and
Superkamiokande) have beautifully demonstrated the capability of such
detectors to study
neutrino emission by the Sun, to investigate the time of flight 
dependence
of neutrino
flavor transitions, to search for baryon number violation via the 
decay of
the proton and
to observe neutrinos from supernova. The largest of these detectors,
Superkamiokande (SK),
has a total water volume of ${\rm 50,000 m^{3}}$ and is at a depth of
approximately 2700 mwe.

We visualize an array of at least ten  detectors for the Homestake
Laboratory, each of 100 kilotons (50m $\times$ 50m $\times$ 50m in
volume), with a total detector mass of the order of a megaton that can
ultimately extend the range of proton decay lifetime searches by two
orders of magnitude from the present value and simultaneously serve as
a hodoscope for ultrahigh energy (UHE) neutrino detection.
The array would provide a high statistics observation of the main 
neutrino emission from a supernova but one of the array filled  with 
a low threshold material for neutrino interactions, e.g., $^{37}Cl$, 
Num
would make possible observation of the $\nu_e$ from 
$e^- + p \rightarrow n + \nu_e$ at the formation of the 
protoneutron star. In addition, the array would serve as a 
target to provide high statistics observations of reactions 
produced by intense sources of neutrinos (e.g. neutrino factories) 
for searches for CP violation in the neutrino sector.

 Locating
the array at a depth of, say, 6000 mwe would result in a cosmic ray
muon flux background per module that is more than two orders of
magnitude less than that at the SK detector and a total muon
background for the entire megaton array that is an order of magnitude
lower than the background in the 50 kton SK detector.
Considering the successes of SK and SNO (at 2700 and 6200 mwe, 
respectively) in coping with
the cosmic ray background at their respective sites, the possibility
exists that the water cerenkov detectors in the Homestake laboratory
might be operated without any, or at least a minimal, loss of useful
volume occupied by all-enclosing veto-counters as has been
conventional
since K-II. This idea might be tested by a ``shotcrete'' coveriNumber of
the surface of the first large room, over which would be an adequate
waterproof coat that would allow the room to be completely filled with
water. The usual array of photomultiplier tubes (PMT) would be
suspended as in the SK detector, saving money and labor, and
permitting
a megaton detector system to be achieved.

After half a century, the deeper questions addressed by elementary 
particle
 physicists---although
broader and subtler---are no less difficult to answer, no less 
directed at
fundamental
issues than the questions they asked when the science began. 
One of the most
basic,
because its ramifications are so extensive, concerns the stability 
of the
proton
against spontaneous decay. Another fundamental question concerns 
the
intrinsic nature
of the neutrino, made even more interesting by the recent observation 
that
neutrinos of
different type possess small but different  mass and are capable of 
making a transition from one
type to
another.

Apart from the profound physics that these questions have in common, 
they
are also
related by the practical need to extend experimental methods beyond 
their
present
boundaries in order to address them. This means, on the one hand, 
extending
the
present lower limit on the proton lifetime beyond its approximate 
value of
$10^{33}$
years and, on the other hand, searching for neutrino behavior in 
phenomena
where new
neutrino properties might be exhibited, for example, at neutrino 
energies
far higher
than any so far observed. Modern techniques indicate that a common 
detector
method
can be employed in both studies provided that the detector can be 
adequately
shielded
against extraneous phenomena, i.e., cosmic rays and their products. 
This has
been
achieved in the past two decades during which the lower limit above 
on the proton
lifetime
was acquired and the science of neutrino astrophysics successfully 
launched.

To go further means appreciably larger, more sophisticated detectors 
and---particularly
important---significantly improved protection of them against cosmic 
rays.
This latter can be
attained by embedding the detectors deeper in the Earth than 
heretofore or
deep under
ice, say at the South Pole, or deep in the sea.  A new
underground
laboratory with the features likely to be found in the Homestake
Laboratory opens that possibility. And the array of detectors 
described above and shown schematically in Fig. 1 allows for 
all of the detectors to be used in an additive mode, i.e., 
as a megaton detector for most experiments, e.g., proton decay, 
and UHE neutrino searches, supernova observations, and CP 
violation search, while an individual member of the array is 
devoted to a different but related purpose, i.e.,  detection 
of the $\nu_e$ as well as the $\bar{\nu}_e$ from a supernova.

{\bf Proton Decay}

As indicated above, the size of the cavity or room module for the
detector that we suggest is about 100 kilotons of ${\rm H_{2}O}$. We
envision building and completing one module as soon as possible after
formation of the underground laboratory with a cylindrical
water-filled Cerenkov detector, equipped with photocathode coverage
and fast sensitive electronics adequate to measure the energies of the
charged products of possible products of proton decay. Roughly 20
percent of the total cylindrical area would be sufficient to allow a
number of the decay modes to be seen; observation and study of
neutrino interactions in the module would make clear necessary
improvements in photocathode coverage and indicate the decay modes for
which the module detection efficiency is especially low.

After the design of the first module and detector are fine-tuned, more
completed module/detector units will be added to the proton-decay
complex at a rate limited by money and cavity excavation progress. A
conservative estimate suggests that a complex of a dozen or so modules
can be excavated and prepared to accept Cerenkov counters within a
period of two to three years for approximately \$150 million. They will
be equipped for particle detection for approximately an equal amount, a
number of them at the same time as excavation of the latter  ones is
proceeding. A possible configuration in which they might be arranged 
is shown in Fig. 1.

Building on the previous work on proton-decay of the earlier water
Cerenkov detectors at Kamioka- Kamiokande,
Kamiokande-II, Superkamiokande, and at
 the Fairport Salt Mine under Lake Erie, few new problems in
analyzing the data from the Homestake complex should arise, and summing the
output of the Homestake detectors should yield an ultimate lower limit on
restricted decay modes of proton-decay about two  orders of magnitude lower
than the present SK limit ($10^{33}$yr).

{\bf UHE Neutrinos}

There are several serious predictions of the possible flux of UHE
neutrinos from various astrophysical and cosmological remnant
sources. The predictions cover a wide range of neutrino energies and
fluxes from a few TeV for the known atmospheric neutrinos to roughly
$10^9$ TeV for neutrinos suggested by speculations on topological
defects; they are shown in Fig. 2 [1].
A clear positive result in the search for UHE neutrinos would open new
unexplored areas in physics, astronomy and cosmology. The modular
detector described above is well-suited for such a search. Roughly, we
expect a large UHE muon-background-free region in the angular interval
$70^0 < \theta < 120^0$, where $\theta$ is the zenith angle defined by the
normal to the plane of the detectors in Fig.1. A UHE neutrino-induced
muon signal, for example, will traverse between one and three modules
depending on their spatial configuration, which will act as a sampling
detector to determine muon direction and muon energy within an order
of magnitude or better. We expect the detector in Fig. 1, depending
on final dimensions, to subtend a solid angle relative to the solid
angle subtended by the canonical ${\rm km^3}$ detector often referred to
in possible UHE neutrino searches of $0.01 < \Delta \Omega <
0.10$. That relative solid angle interval would be satisfactory for an
initial survey in which the goal would be to observe a small number of
events to show that UHE neutrinos exist. A positive signal from the
proposed detector or any other would be impetus to expand the array in
Fig. 1 further.

 It is realistic to consider
the prospect of observing  low intensity fluxes such as shown in Fig. 2 
because the
reaction cross section for neutrino plus nucleon rises steeply with
increasing neutrino energy for both neutrinos and antineutrinos. For
example, the cross section rises by five decades as the neutrino
energy increases by seven decades despite the production of real
intermediate vector bosons. This is given in more detail in Fig. 3 [2].

An important consequence of the rapid increase of neutrino interaction
cross sections is that the Earth becomes a significant absorber of
neutrinos above roughly $10^4$ GeV and is essentially opaque to
neutrinos with energy above $10^7$ GeV. The neutrino survival
probability as a function of $\cos \theta_Z$ for those neutrino
energies
is plotted in Fig. 4; $\theta_Z$ is the zenith angle at the
detector [3]. In obtaining the curves in Fig. 4, the Earth is modeled as
a high density  ($ {\rm 15 gm/cm^3}$) core and a low density mantle,
which accounts for the sharp break near large negative values of $\cos
\theta_Z$ at the lower energies. This effect requires serious
consideration in 
formulating plans for UHE neutrino searches, especially if the actual
fluxes are much lower than suggested by Fig. 1. 

A less dramatic, but nevertheless highly interesting, measurement that 
also becomes possible with the array and the increase of neutrino cross 
sections is the mapping of the density of the Earth's interior as described 
briefly in Fig. 5 [4].

{\bf Supernovae Neutrinos}

Since the first observations of neutrinos from the type II supernova, 
SN 1987A, the intent of neutrino astrophysicists is to be prepared to 
study the neutrino emission from a type II supernova in our own or a 
nearby galaxy if and when the occasion arises.  One of the detectors 
in the array in Fig. 1 would provide a good statistical sample of such 
neutrinos for analysis; however, utilization
of another member of the array for a different but correlated measurement 
might be particularly rewarding if the second module is properly equipped.

The current model of a type II supernova core collapse involves a step in 
which the protoneutron star is formed through complete dissociation of the 
components of the nuclei in the collapsed iron core followed by a newly 
freed proton  capturing an electron;  the reaction giving rise to a neutron 
and  $\nu_e$. 
The resulting burst of $\nu_e$ was not observed during SN 1987A because 
it is expected to be short-lived ($\leq 1 sec$), and to carry only about 
ten percent of the total neutrino energy emitted by the supernova in the 
subsequent $\nu_e$,
$\bar{\nu}_e$, etc. pairs. Measurement of the short $\nu_e$ pulse would 
mark the instant from which to measure all later times of interest in the 
event.

However, in a water Cerenkov detector, the low energy of $\bar{\nu}_e$ 
from a supernova are much easier to detect than the $\nu_e$ because the 
two protons in the water molecule are essentially free to participate 
in the reaction $\bar{\nu}_e + p \rightarrow e^- + n$, while the 
neutrons which are required to satisfy separate lepton number and 
charge conservation in the particle reaction 
$\nu_e + n \rightarrow e^- +p$ are tightly bound in the oxygen 
nucleus and accordingly demand higher energy neutrinos to make the 
latter particle reaction go. The $\bar{\nu}_e$ from SN 1987A were 
in fact the ones detected.

If a second module contained chlorine in the form, say of $NaCl$, 
however, the reaction $\nu_e + ^{37} Cl \rightarrow e^- + ^{37}Ar$ 
would occur in it
 with a neutrino energy threshold of 0.814 MeV; and electrons with 
kinetic energy above roughly 7 MeV from that reaction would be 
detected as would their emission times and energies. This is shown 
schematically in Figure 6. A comparison of the energy and time spectra 
of the $\nu_e^-$ and $\bar{\nu}_e^-$ induced electron spectra from the 
two reactions would describe the direct neutrino emission from the 
supernova completely with at least adequate statistical samples.
Diversion of one of the 50 kiloton detectors for this purpose would 
not render it unuseful for the other experiments described here.

{\bf Search for CP Violation in the Neutrino Sector}

Still another interesting use of the massive, multipurpose neutrino 
detectors is to search for CP violation in the neutrino sector with 
an intense, accelerator-produced neutrino beam. The principle involved 
in the search is a precision comparison of neutrino reactions in which 
the rates are required by CP invariance to be equal. 

The neutrino detector array in the Homestake Laboratory is particularly 
well-suited to conduct a search for CP-invariance violation in the 
neutrino sector because of:

\begin{enumerate}
\item[] (a) its large mass to provide high reaction rates when combined 
with high neutrino fluxes

\item[] (b) its high sensitivity to charged leptons over a wide momentum 
range and its capability to measure accurately momentum and direction of 
the lepton

\item[] (c) with different fillings (as in the supernova measurement above),
the array can distinguish between $\nu_x$ and $\bar{\nu}_x$ at low $\nu$ 
energy

\item[] (d) the distance at which the CP violating effect is likely to 
appear is large, i.e.\\
\noindent  $\geq$ 3000 km., so that an intense accelerator neutrino beam 
or neutrino factory 
beam can be aimed at the Homestake Laboratory and be effective.
\end{enumerate}

{\bf Summary and Conclusions}

This brief note outlines a portion of the physics experiments that 
might be carried out with the megaton modular multi-purpose neutrino 
detector array described herein. The physical properties of the 
Homestake Mine are especially favorable for the construction and 
operation of a laboratory to accommodate such a detector array which 
would be difficult or impossible to house in most other
deep mines. These experiments are among the most penetrating to be 
done in the neutrino sector and their results will contribute to the 
resurgence of that sector in astro-particle physics.

{\bf Acknowledgement}

This note is the product of many discussions with Ken Lande and 
information about Homestake obtained from him. 

{\bf References}

\begin{enumerate}
\item L. Nellen, K. Mannheim, P.L. Biermann, Phys. Rev. D 47 (1993) 5270; 
A.P. SZabo, R.J. Protheroe, Astropart. Phys. 2 (1994) 375; K. Mannheim, 
Astropart. Phys. 3 (1995) 295;
D. Kazanas, in: Proc. Third NESTOR International Workshop, October 1993, 
L.K. Resvanis, ed. (Athens Univ. Press, 1994); F.W. Stecker, M.H. Salamon, 
Space Sci. Rev. 75 (1996) 341; E. Waxman, J. Bahcall, 
Phys. Rev. Lett. 78 (1997) 2292; Phys. Rev. D 59 (1999) 023002.

\item R. Gandhi, C. Quigg, M.H. Reno, I. Sarcevic, 
Astropart. Phys. 5 (1996) 81; G.C. Hill, Astropart. Phys. 6 (1997) 215.

\item Astropart Physics 10, 321 (1999).

\item Alfred K. Mann, Proc. Seventh Int'l Workshop on Neutrino Telescopes, 
Venice, 1996, ed. Milla Baldo Ceolin, p. 415. 
\end{enumerate}

\end{document}